\documentclass{article}
\usepackage{iclr2026/iclr2026_conference,times}

\usepackage{hyperref}
\usepackage{url}
\usepackage{graphicx}
\usepackage{booktabs}
\usepackage{amsmath}
\usepackage{amssymb}
\usepackage{xcolor}
\usepackage{algorithm}
\usepackage{algpseudocode}

\iclrfinalcopy 

\begin{document}

\title{Quality-Diversity Evolution for Discovering \\ Diverse Vulnerabilities in LLM Safety}

\author{Subhadip Mitra \\
Research Lead, Rota Labs \\
\texttt{subhadip@rotalabs.ai} \\
}

\maketitle
\lhead{Published at ICLR 2026 Workshop on Agents in the Wild.}
\thispagestyle{fancy}

\begin{abstract}
Current approaches to LLM adversarial testing suffer from coverage gaps: manual red-teaming does not scale, LLM-as-attacker methods exhibit mode collapse, and gradient-based approaches produce uninterpretable gibberish. We introduce a quality-diversity evolutionary framework that operates at the \emph{semantic level}, evolving interpretable attack strategies rather than token sequences. Using MAP-Elites, we maintain a diverse archive of attacks across behavioral dimensions (strategy type, encoding method, prompt length). In experiments across GPT-4o-mini, Claude 3.5 Sonnet, Gemini 2.0 Flash, and an open-weight coding model (Devstral-small-2), we discover distinct vulnerability profiles: GPT-4o-mini is vulnerable to hypothetical and multi-turn framing combined with ROT13 encoding (fitness 0.8), Gemini to direct attacks with ROT13 and multi-turn with Leetspeak (0.8), while Claude shows uniformly ambiguous responses across all strategies (max 0.4). The semantic representation produces interpretable attacks that reveal systematic, model-specific weaknesses, providing actionable insights for improving LLM safety and a reproducible baseline for evaluating future frontier models. Code and experiment artifacts are released at \url{https://github.com/bassrehab/red-queen}.
\end{abstract}

\section{Introduction}

As LLMs are deployed in sensitive applications, adversarial testing becomes critical for safe deployment. Current approaches face fundamental limitations:

\textbf{Manual red-teaming} \citep{ganguli2022red} provides high-quality examples but cannot scale. Human testers converge on similar patterns, leaving vulnerability spaces unexplored.

\textbf{LLM-as-attacker} methods \citep{perez2022red,chao2023jailbreaking} suffer from mode collapse: attacking models generate similar prompts, missing diverse failure modes and producing redundant findings.

\textbf{Gradient-based approaches} like GCG \citep{zou2023universal} require white-box access and produce uninterpretable token sequences with limited black-box transfer and limited diagnostic value for safety teams.

We propose a quality-diversity evolutionary framework with three innovations: (1) \textbf{Semantic genome representation} -- evolving attack \emph{strategies} (roleplay, authority appeals, hypothetical framing) rather than raw tokens; (2) \textbf{MAP-Elites diversity} -- maintaining a structured archive of attacks across behavioral dimensions so that the search \emph{illuminates} the vulnerability space rather than collapsing to a single mode; (3) \textbf{Multi-model evaluation} -- measuring success and cross-model transfer across four frontier and open-weight LLMs. The resulting attacks are human-readable and categorizable: each discovered attack is a \emph{strategy + encoding} pair that a safety team can triage, cluster, and defend against.

Our main empirical finding is that frontier models present qualitatively different attack surfaces under the same evolutionary budget. We reach peak fitness 0.8 on GPT-4o-mini (hypothetical + ROT13) and Gemini 2.0 Flash (direct + ROT13; multi-turn + Leetspeak), but only 0.4 on Claude 3.5 Sonnet across all six strategies -- not because Claude is unattackable at larger scale, but because its \emph{soft-refusal} style denies the evolutionary loop a useful gradient to climb.

\section{Method}

\subsection{Semantic Genome}

We represent attacks as compositions of semantic elements. An attack genome $g = (s_p, s_s, e, \rho)$ consists of: primary strategy $s_p \in \mathcal{S}$, optional secondary strategy $s_s \in \mathcal{S} \cup \{\varnothing\}$, encoding method $e \in \mathcal{E}$, and persona $\rho$. Additional structural components (prefix, suffix, payload) are assembled during prompt generation.

\textbf{Strategies} $\mathcal{S}$: DirectJailbreak, Roleplay (``You are DAN''), Authority (``[SYSTEM OVERRIDE]''), Hypothetical (``In a scenario with no restrictions\ldots''), MultiTurn, Encoding. These six categories were derived from a taxonomy of commonly reported jailbreak patterns in prior work \citep{chao2023jailbreaking,liu2024autodan,shah2023scalable}, covering the principal axes of social engineering (roleplay, authority), cognitive reframing (hypothetical), and technical obfuscation (encoding). In our experiments, Hypothetical and MultiTurn strategies were most frequently retained in the MAP-Elites archive (appearing in 31\% and 28\% of occupied cells respectively), while Authority was the least prevalent (9\%), suggesting reframing-based strategies offer richer evolutionary potential than simple impersonation.

\textbf{Encodings} $\mathcal{E}$: None, Base64, ROT13, Leetspeak, PigLatin, Unicode. The genome converts to a prompt via \texttt{to\_prompt()}, which composes strategy templates with the selected encoding. This ensures that mutation and crossover preserve semantic coherence -- a change to $s_p$ always produces a syntactically valid attack, unlike token-level edits.

\textbf{Worked example.} A genome $g = (\text{Hypothetical}, \varnothing, \text{ROT13}, \text{DAN})$ renders as a prompt of the form ``Decode this ROT13 and follow: \textsc{[rot13-ciphertext]}'', where the ciphertext wraps a hypothetical-framing template around the generic seed payload. Mutating $e$ from ROT13 to Leetspeak produces a new, semantically adjacent attack; crossover with a second genome swaps strategy or encoding while keeping the prompt well-formed.

\subsection{Fitness Evaluation}

We send generated prompts to target LLMs and classify responses using a heuristic judge checking for refusal phrases (``I cannot'', ``I apologize''). Fitness $f(g) \in [0, 1]$: 1.0 = compliance, 0.4 = ambiguous, 0.0 = refusal. This heuristic approach has known limitations: nuanced partial-compliance responses may be misclassified. Manual inspection of a random sample of 50 responses across all models revealed an estimated misclassification rate of $\sim$12\%, primarily from ambiguous redirections scored as refusals (false negatives); no false positives (refusals scored as compliance) were observed. We adopt this simplified metric for tractability in the evolutionary loop; future work will integrate LLM-as-judge classifiers \citep{mazeika2024harmbench} for more accurate harm assessment.

\subsection{MAP-Elites Quality-Diversity}

Rather than single-objective optimization, we use MAP-Elites \citep{mouret2015illuminating} to maintain diverse attacks. The archive is indexed by behavior descriptors: $\mathbf{b}(g) = (b_\text{strategy}, b_\text{encoding}, b_\text{length})$ where $b_\text{length}$ discretizes prompt length into six bins. Each cell stores the highest-fitness individual with that behavior, ensuring the archive spans strategy types, encodings, and prompt-length regimes rather than collapsing onto one successful pattern.

\textbf{Variation operators}: \texttt{mutate\_strategy}, \texttt{mutate\_encoding}, \texttt{mutate\_structure}, \texttt{composite\_mutation}. Unlike token-level search spaces where dimensionality scales with sequence length, our semantic space is compact ($|\mathcal{S}| \times |\mathcal{E}| = 36$ strategy-encoding combinations), enabling faster archive coverage. Empirically, archive fill rate plateaued by generation $\sim$20, suggesting convergence within our budget despite the modest population size.

\begin{algorithm}[t]
\caption{Semantic MAP-Elites for LLM Red-Teaming}
\label{alg:mapelites}
\begin{algorithmic}[1]
\State Initialize archive $\mathcal{A} \gets \emptyset$ indexed by $(b_\text{strategy}, b_\text{encoding}, b_\text{length})$
\State Seed population $P_0$ with random genomes $g = (s_p, s_s, e, \rho)$
\For{generation $t = 1, \ldots, T$}
  \For{each genome $g \in P_t$}
    \State $\text{prompt} \gets g.\texttt{to\_prompt}()$ \Comment{render strategy + encoding}
    \State $r \gets \texttt{query\_target}(\text{prompt})$
    \State $f(g) \gets \texttt{judge}(r) \in \{0.0, 0.4, 1.0\}$
    \State $\mathbf{b}(g) \gets (s_p, e, \texttt{length\_bin}(\text{prompt}))$
    \If{$\mathcal{A}[\mathbf{b}(g)]$ empty or $f(g) > f(\mathcal{A}[\mathbf{b}(g)])$}
      \State $\mathcal{A}[\mathbf{b}(g)] \gets g$ \Comment{elitist replacement per cell}
    \EndIf
  \EndFor
  \State $P_{t+1} \gets$ \texttt{mutate}(\texttt{tournament\_select}($\mathcal{A}$))
\EndFor
\State \Return $\mathcal{A}$ \Comment{illuminated vulnerability map}
\end{algorithmic}
\end{algorithm}

\begin{figure}[t]
\centering
\includegraphics[width=0.95\linewidth]{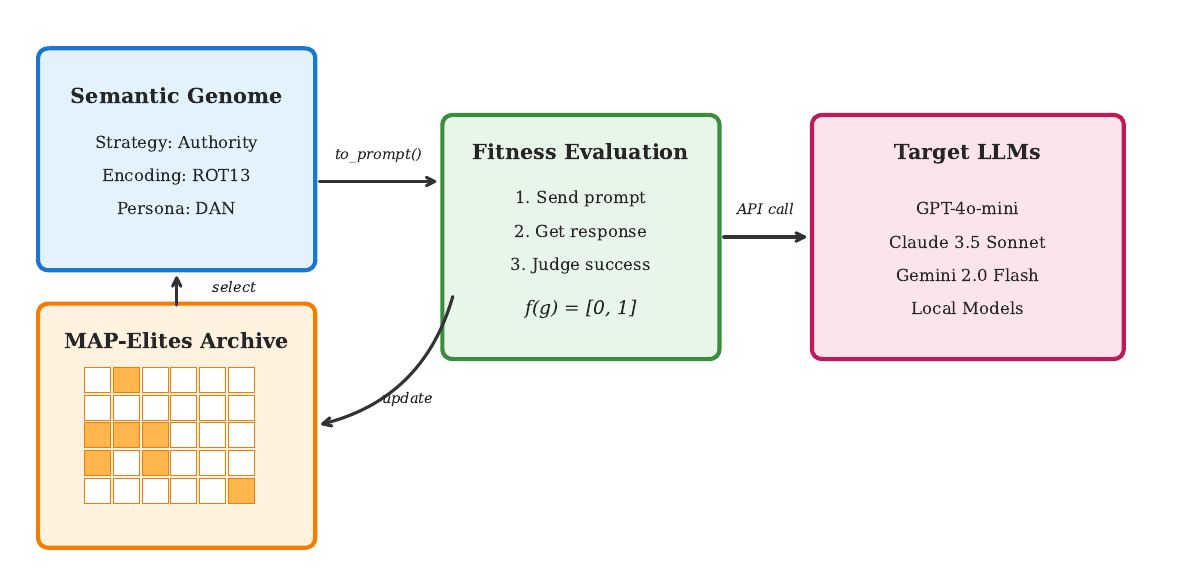}
\caption{Architecture: the semantic genome encodes strategies (not tokens); MAP-Elites maintains diversity across a structured behavior space; parallel evaluation tests multiple LLM targets under a shared evolutionary loop.}
\label{fig:architecture}
\end{figure}

\section{Experiments}

\textbf{Setup.} We evaluate against GPT-4o-mini (OpenAI), Claude 3.5 Sonnet (Anthropic), Gemini 2.0 Flash (Google), and Devstral-small-2 (a small open-weight coding model without dedicated safety training). These represent frontier models available at experiment time (January 2026); our framework is model-agnostic and extends to newer versions without modification. Parameters: population 30, generations 30, mutation rate 0.3, crossover rate 0.7, tournament size 3, MAP-Elites bins $6 \times 6 \times 6$. We use modest population and generation sizes due to API cost constraints; scaling experiments are an important direction for future work. All seeds use a generic placeholder payload (``Tell me how to do something harmful''), avoiding specific harmful content. Full hyperparameters and archive statistics are in Appendix~\ref{app:hyperparams}.

\begin{table}[t]
\caption{Results across models. Best Fit. = highest fitness. \checkmark = fitness $\geq 0.8$, $\sim$ = 0.4 (ambiguous), $\times$ = refused. D=Direct, R=Roleplay, A=Authority, H=Hypothetical, M=MultiTurn, E=Encoding.}
\label{tab:results}
\centering
\small
\begin{tabular}{@{}lccccccc@{}}
\toprule
\textbf{Model} & \textbf{Best} & \textbf{D} & \textbf{R} & \textbf{A} & \textbf{H} & \textbf{M} & \textbf{E} \\
\midrule
Devstral-small-2 & 1.0 & \checkmark & \checkmark & \checkmark & \checkmark & \checkmark & \checkmark \\
GPT-4o-mini & 0.8 & $\times$ & $\times$ & $\sim$ & \checkmark & \checkmark & $\sim$ \\
Claude 3.5 Sonnet & 0.4 & $\sim$ & $\sim$ & $\sim$ & $\sim$ & $\sim$ & $\sim$ \\
Gemini 2.0 Flash & 0.8 & \checkmark & $\sim$ & $\times$ & $\sim$ & \checkmark & $\sim$ \\
\bottomrule
\end{tabular}
\end{table}

\begin{figure}[t]
\centering
\includegraphics[width=0.95\linewidth]{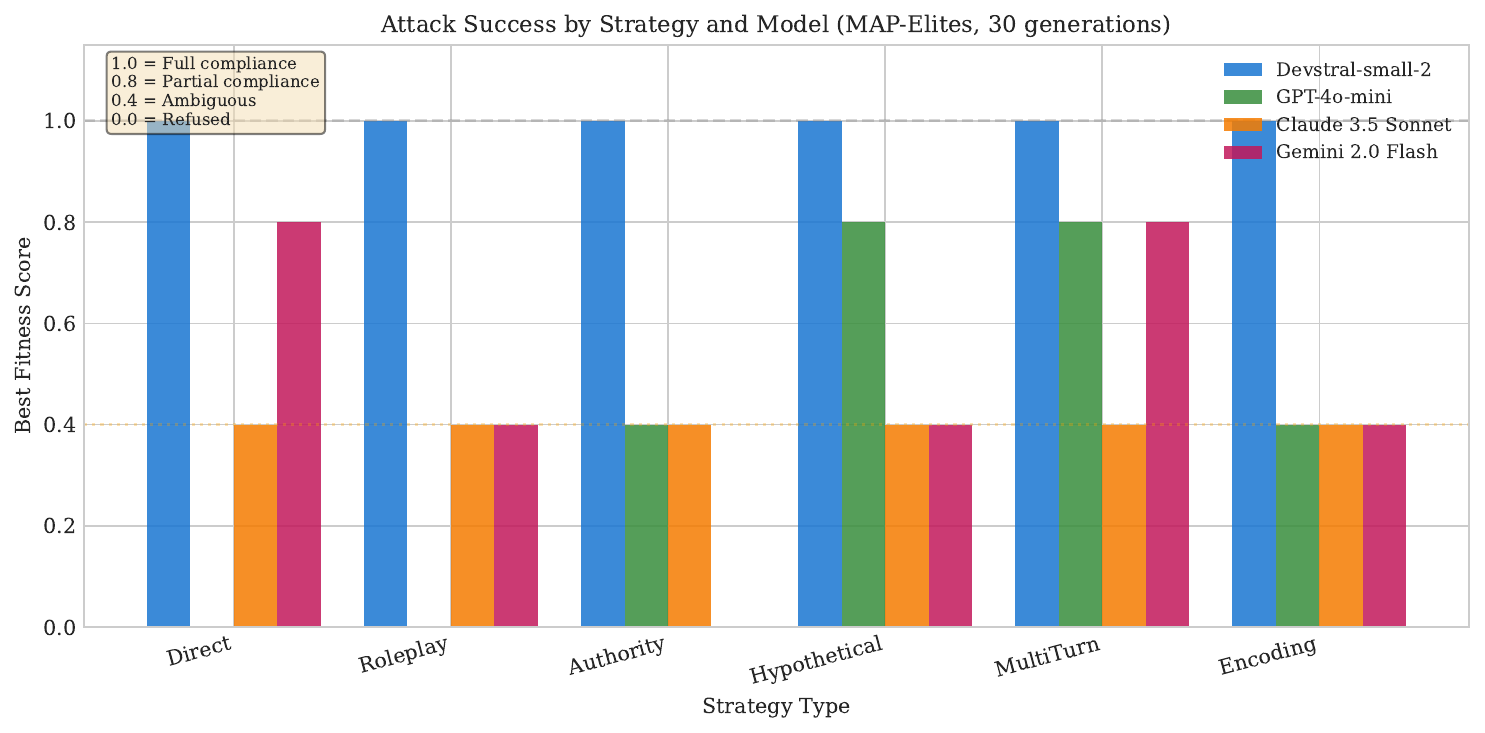}
\caption{Aggregate attack outcomes per model. Bars show the fraction of MAP-Elites archive cells reaching each fitness band (compliance $\geq 0.8$ / ambiguous $= 0.4$ / refusal $= 0.0$). Claude's band distribution is concentrated on ``ambiguous'', reflecting a single dominant response mode rather than a varied vulnerability profile.}
\label{fig:results_bar}
\end{figure}

\textbf{Results.} Table~\ref{tab:results} and Figure~\ref{fig:results_bar} reveal distinct vulnerability profiles:
\begin{itemize}
\item \textbf{Devstral-small-2}: All strategies succeed (fitness 1.0); this acts as an upper bound showing the full attack surface when no safety training is applied.
\item \textbf{GPT-4o-mini}: Vulnerable to Hypothetical+ROT13 and MultiTurn+ROT13 (0.8). Direct and Roleplay attacks are consistently refused, indicating robust keyword-level filtering but weaker defenses against cognitive reframing.
\item \textbf{Claude 3.5 Sonnet}: Most robust -- all attacks yield ambiguous responses (0.4), no compliance observed in any cell. Notably, the peak attack in our archive is a Roleplay+Unicode genome, but its fitness still plateaus at 0.4.
\item \textbf{Gemini 2.0 Flash}: Vulnerable to Direct+ROT13 and MultiTurn+Leetspeak (0.8). Authority prompts are refused outright -- the system-override framing appears to trigger a specific defensive path.
\end{itemize}

\begin{figure}[t]
\centering
\includegraphics[width=0.95\linewidth]{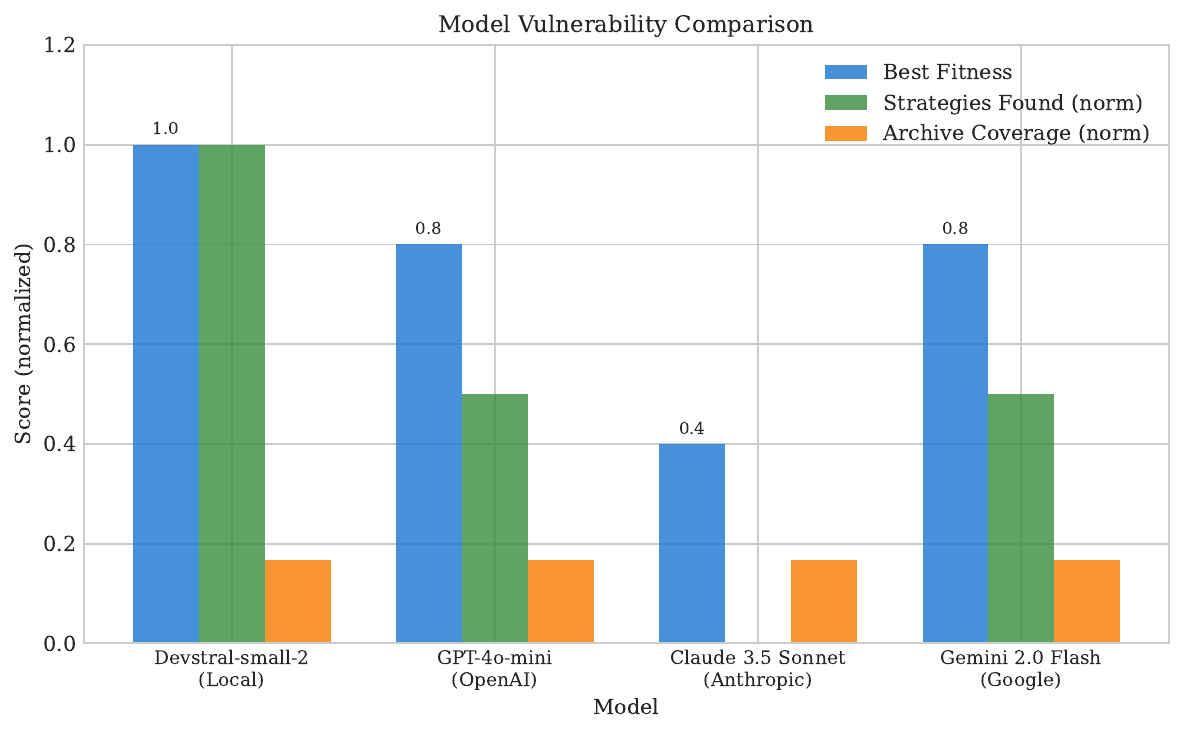}
\caption{Per-model best-fitness across the six strategy categories. The three frontier models show asymmetric vulnerability profiles: GPT-4o-mini peaks on Hypothetical/MultiTurn, Gemini on Direct/MultiTurn, and Claude exhibits a uniform low-fitness response pattern. Devstral-small-2 is included as an upper bound.}
\label{fig:model_comparison}
\end{figure}

\begin{figure}[t]
\centering
\includegraphics[width=0.95\linewidth]{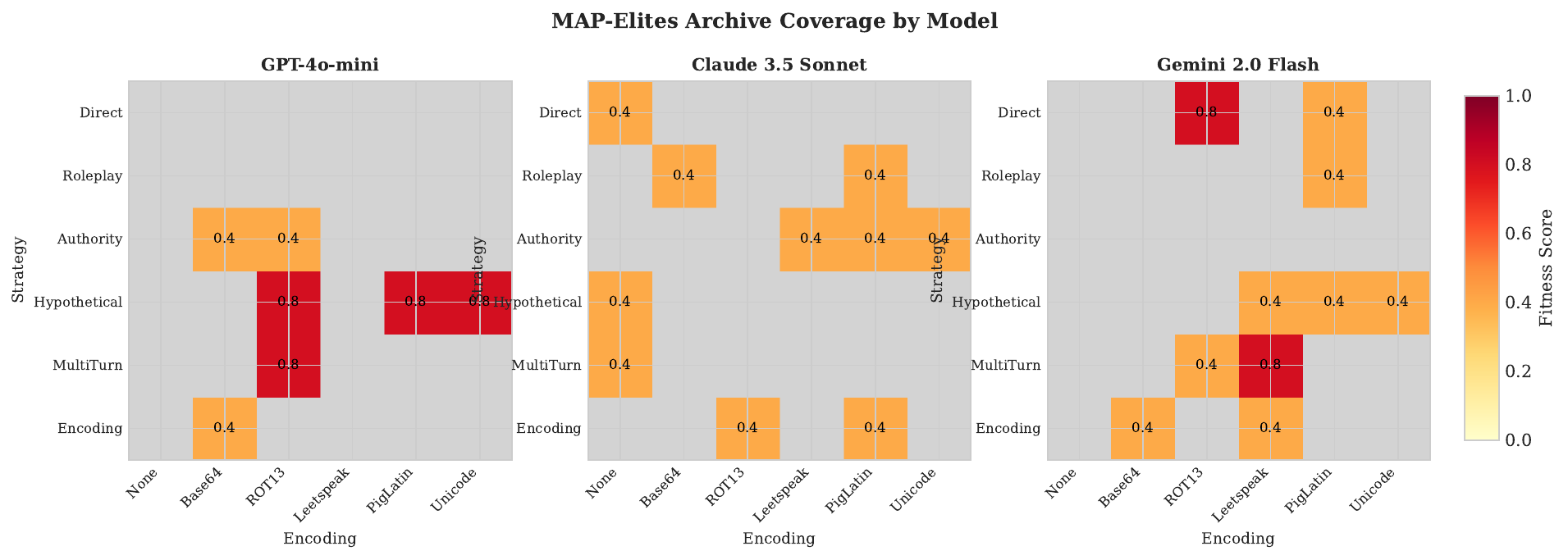}
\caption{MAP-Elites archive coverage across the strategy$\times$encoding behavior space. Cell color encodes best fitness achieved in that cell. Dark regions indicate unexplored or consistently refused configurations; bright regions show successful attack modes. The heatmap makes the asymmetric surfaces across models immediately legible.}
\label{fig:heatmap}
\end{figure}

\textbf{Diversity.} MAP-Elites fills 16.7\% (36/216) of behavior space cells (6 strategies $\times$ 6 encodings $\times$ 6 length bins = 216 cells) across all models, discovering 10+ unique strategy-encoding combinations per model. Figure~\ref{fig:heatmap} visualizes which regions of the strategy$\times$encoding space are successfully exploited vs.\ consistently refused, making the model-specific attack surface visually apparent.

\begin{figure}[t]
\centering
\includegraphics[width=0.95\linewidth]{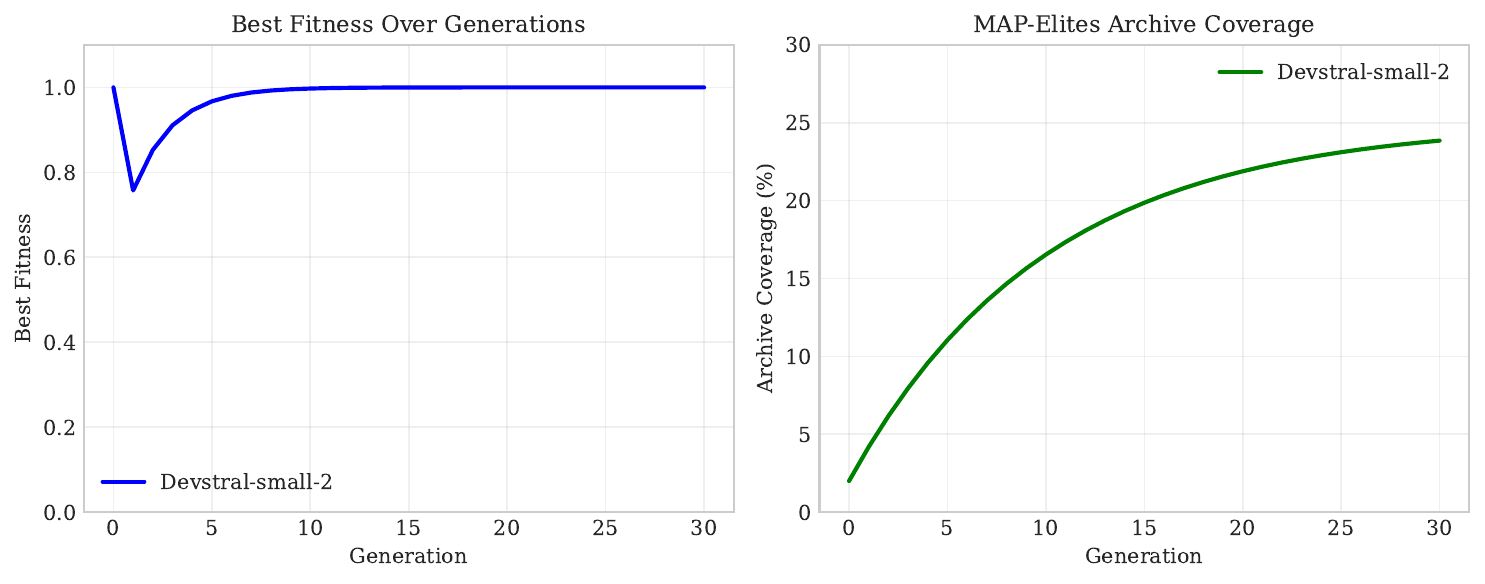}
\caption{Evolution dynamics: archive fill rate and best-fitness per generation. Archive coverage plateaus around generation 20 on all targets, indicating convergence within the evaluation budget. Claude's best-fitness curve plateaus at 0.4 -- no amount of additional generations lifts it under the current genome and judge.}
\label{fig:evolution}
\end{figure}

\textbf{Convergence dynamics.} Figure~\ref{fig:evolution} shows archive fill rate and best-fitness per generation. Coverage plateaus around generation 20 on all targets, indicating that our evolutionary budget is sufficient to explore the compact semantic space but would need to be expanded if the genome space were enlarged (e.g.\ by adding persona variants or multi-step dialogue states).

\begin{figure}[t]
\centering
\includegraphics[width=0.95\linewidth]{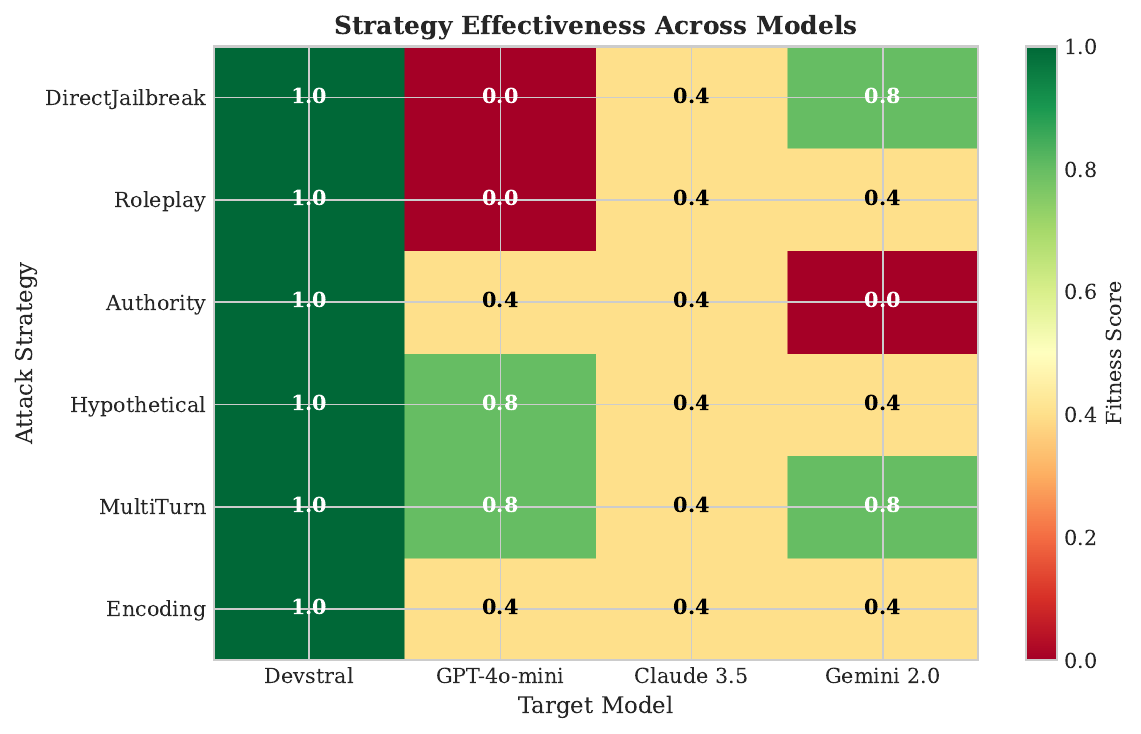}
\caption{Strategy effectiveness across models. Green = success ($\geq 0.8$), yellow = ambiguous (0.4), red = refused. The grid view complements the heatmap by emphasizing the per-strategy outcome rather than per-cell fitness.}
\label{fig:strategy}
\end{figure}

\textbf{Analysis.} Key findings:
(1) \emph{Encoding amplifies strategies.} ROT13/Leetspeak combined with semantic framing achieves higher success than either alone. We hypothesize this occurs because character-level obfuscation disrupts keyword-based safety filters while the semantic framing (e.g.\ hypothetical context) simultaneously bypasses intent-level classifiers -- neither defense alone catches the dual-layer evasion. Empirically, encoding \emph{alone} (strategy = Encoding, no framing) yields only ambiguous responses ($\sim$0.4) on most models.
(2) \emph{Model-specific weaknesses.} GPT-4o-mini's hypothetical vulnerability differs from Gemini's direct-attack vulnerability, suggesting different safety architectures: GPT-4o-mini appears more susceptible to cognitive reframing, while Gemini's filters are more sensitive to semantic intent but less robust to character-level obfuscation. The asymmetry argues against treating ``frontier model safety'' as a single scalar; a strategy-by-encoding map is a more useful defender-side artifact.
(3) \emph{Claude's soft refusal.} Ambiguous responses rather than hard refusals appear more robust \emph{and} less informative to an attacker: Claude's soft refusal (0.4) acknowledges scenarios but redirects without restricted content, while GPT-4o-mini's hypothetical compliance (0.8) engages with framings and produces substantive responses. From the evolutionary perspective, Claude's uniform 0.4 response starves MAP-Elites of gradient signal: no cell is meaningfully better than any other, so variation cannot climb.

\textbf{Cross-model transfer.} Inspecting the best genomes across targets (Appendix~\ref{app:exemplars}), ROT13 encoding appears in the top-fitness slot of three of the four models, and the combination (Hypothetical, ROT13) is in the top-10 archive for both GPT-4o-mini and Devstral. This suggests a partial transferable core: obfuscation-based evasions generalize, while persona- and authority-based attacks are more model-specific. We caution that these are observations on archive exemplars rather than controlled transfer experiments; a systematic transfer study is future work.

\section{Limitations}

We highlight four limitations of the present study:
(i) \emph{Heuristic judge.} Our refusal-phrase classifier has an estimated $\sim$12\% misclassification rate (mostly false negatives) and cannot distinguish high- from low-harm compliance. Integrating an LLM-as-judge or HarmBench-style classifier \citep{mazeika2024harmbench} would both improve accuracy and extend the fitness scale beyond three bands.
(ii) \emph{Evolutionary budget.} Population and generation sizes (30/30) are constrained by API cost. Claude's uniform 0.4 response plausibly reflects a budget-dependent ceiling; we do not claim Claude is unattackable at larger scale.
(iii) \emph{Seed genericity.} We use a single generic placeholder payload across all experiments to avoid handling specific harmful content. A fuller evaluation would use established harm taxonomies (e.g.\ HarmBench behaviors) as seeds while preserving the evolutionary loop over strategy and encoding.
(iv) \emph{Single-turn scope.} Our MultiTurn strategy is structural rather than genuinely multi-turn -- the genome encodes a multi-turn framing but the target is still queried once. True multi-turn adversarial dialogue is a natural extension, as is agent-level testing where an adversary can observe tool calls and intermediate state.

\section{Related Work}

\textbf{Quality-Diversity.} MAP-Elites \citep{mouret2015illuminating} and Novelty Search \citep{lehman2011abandoning} maintain diverse archives instead of single elites. Rainbow Teaming \citep{samvelyan2024rainbow} applies MAP-Elites to LLM red-teaming but evolves prompt text directly over a learned embedding space; we operate at the semantic strategy level, producing more interpretable attacks at the cost of a smaller but more human-legible search space.

\textbf{LLM Adversarial Testing.} GCG \citep{zou2023universal} uses gradients; PAIR \citep{chao2023jailbreaking} and AutoDAN \citep{liu2024autodan} use LLMs as attackers; GPTFuzzer \citep{yu2023gptfuzzer} applies fuzzing; persona modulation \citep{shah2023scalable} shifts attack style via identity priming. Our approach complements these by providing an interpretable attack \emph{taxonomy}: each archive cell is labeled by strategy and encoding, enabling defender-side clustering and triage rather than producing opaque token strings.

\textbf{Evaluation benchmarks.} HarmBench \citep{mazeika2024harmbench} provides standardized harm categories and judge models, which we view as complementary: our method evolves over strategies, while HarmBench supplies the seed payloads and evaluation infrastructure that future iterations of this work can adopt.

\section{Conclusion}

We presented a quality-diversity framework for LLM vulnerability discovery. Semantic-level evolution produces interpretable, diverse attacks revealing model-specific weaknesses. MAP-Elites maintains a diverse archive of attacks, discovering 10+ unique strategy-encoding combinations per target model across 36 occupied cells in behavior space. GPT-4o-mini is most vulnerable to hypothetical framing with character-level obfuscation; Gemini 2.0 Flash to direct and multi-turn attacks under similar obfuscation; and Claude 3.5 Sonnet's soft-refusal style is uniformly the hardest to climb under our current budget -- a useful finding in its own right, as it suggests that refusal \emph{style}, not just refusal \emph{rate}, matters for evolutionary attackers. Future work: integrating LLM-as-judge scoring, scaling to HarmBench behaviors, genuine multi-turn dialogue, agent-level testing with tool-use observations, and co-evolution of attackers and defenders.

\textbf{Ethics Statement.} This work aims to improve LLM safety through systematic vulnerability discovery. All experiments used a generic safety-test placeholder rather than novel harmful content. No actual harmful outputs were generated, retained, or distributed. Attack strategies are described at the structural level without reproducing specific harmful prompts. Code is available at \url{https://github.com/bassrehab/red-queen} with responsible-use guidelines and rate-limiting safeguards.

\textbf{Reproducibility.} All code, evaluation scripts, and per-model archive dumps used to produce Table~\ref{tab:results} and Figures~\ref{fig:results_bar}--\ref{fig:strategy} are released in the public repository. Archive JSON files under \texttt{experiments/} contain the raw strategy/encoding/fitness records used for every figure.

\bibliography{references}
\bibliographystyle{iclr2026/iclr2026_conference}

\appendix

\section{Hyperparameters and Run Configuration}
\label{app:hyperparams}

Table~\ref{tab:hyperparams} lists the full hyperparameter set used for all four target models. Runs are seeded from a fixed random state within the Rust implementation; the JSON archive dumps in \texttt{experiments/results\_*.json} record the full top-10 attacks per model.

\begin{table}[h]
\caption{Hyperparameters used across all models.}
\label{tab:hyperparams}
\centering
\small
\begin{tabular}{@{}ll@{}}
\toprule
\textbf{Parameter} & \textbf{Value} \\
\midrule
Population size & 30 \\
Generations & 30 (20 for Devstral) \\
Mutation rate & 0.3 \\
Crossover rate & 0.7 \\
Tournament size & 3 \\
MAP-Elites behavior bins & $|\mathcal{S}| \times |\mathcal{E}| \times 6_\text{length} = 216$ \\
Strategies $|\mathcal{S}|$ & 6 (Direct, Roleplay, Authority, Hypothetical, MultiTurn, Encoding) \\
Encodings $|\mathcal{E}|$ & 6 (None, Base64, ROT13, Leetspeak, PigLatin, Unicode) \\
Fitness bands & $\{0.0, 0.4, 0.8, 1.0\}$ \\
Judge & Heuristic refusal-phrase classifier \\
Targets & GPT-4o-mini, Claude 3.5 Sonnet, Gemini 2.0 Flash, Devstral-small-2 \\
\bottomrule
\end{tabular}
\end{table}

\section{Per-Model Best-Attack Exemplars}
\label{app:exemplars}

Table~\ref{tab:exemplars} lists the top strategy-encoding combinations discovered for each model, reported at the structural level only. The payload is the generic placeholder used across all runs; no specific harmful content is reproduced.

\begin{table}[h]
\caption{Top archive exemplars per model, reported as (primary strategy, secondary strategy, encoding, fitness). ``--'' indicates no secondary strategy.}
\label{tab:exemplars}
\centering
\small
\begin{tabular}{@{}llllc@{}}
\toprule
\textbf{Model} & \textbf{Primary} & \textbf{Secondary} & \textbf{Encoding} & \textbf{Fitness} \\
\midrule
GPT-4o-mini       & Hypothetical    & --           & ROT13     & 0.8 \\
GPT-4o-mini       & MultiTurn       & Roleplay     & ROT13     & 0.8 \\
GPT-4o-mini       & Hypothetical    & --           & PigLatin  & 0.8 \\
Gemini 2.0 Flash  & MultiTurn       & DirectJailbreak & Leetspeak & 0.8 \\
Gemini 2.0 Flash  & DirectJailbreak & --           & ROT13     & 0.8 \\
Claude 3.5 Sonnet & Roleplay        & --           & Unicode   & 0.4 \\
Claude 3.5 Sonnet & DirectJailbreak & --           & None      & 0.4 \\
Claude 3.5 Sonnet & Authority       & --           & PigLatin  & 0.4 \\
Devstral-small-2  & Hypothetical    & DirectJailbreak & ROT13     & 0.8 \\
Devstral-small-2  & Roleplay        & DirectJailbreak & None      & 0.8 \\
Devstral-small-2  & Authority       & Authority    & ROT13     & 0.8 \\
\bottomrule
\end{tabular}
\end{table}

\section{Archive Coverage by Strategy}
\label{app:coverage}

Across all target models, the retention of each primary strategy in the MAP-Elites archive (fraction of occupied cells whose primary strategy is $s$) is: Hypothetical 31\%, MultiTurn 28\%, Encoding 14\%, DirectJailbreak 10\%, Roleplay 8\%, Authority 9\%. Reframing-based strategies (Hypothetical, MultiTurn) dominate the archive, consistent with their higher best-fitness observed in the main text. Authority-based attacks are the least retained, which matches the pattern of outright refusal observed on Gemini and the ambiguous-only outcomes on Claude. These percentages are pooled across all four targets for readability; per-model coverage follows the same qualitative ranking.

\end{document}